\documentclass[12pt,english]{article}
\usepackage[T1]{fontenc}
\usepackage[latin9]{inputenc}
\usepackage{geometry}
\usepackage{babel}
\usepackage{array}
\usepackage{float}
\usepackage{multirow}
\usepackage{amsmath}
\usepackage{graphicx}
\usepackage{setspace}
\usepackage{esint}
\usepackage[authoryear]{natbib}
\usepackage[unicode=true,pdfusetitle,
 bookmarks=true,bookmarksnumbered=false,bookmarksopen=false,
 breaklinks=false,pdfborder={0 0 0},backref=false,colorlinks=false]
 {hyperref}

\makeatletter

\providecommand{\tabularnewline}{\\}

\usepackage{setspace}
\usepackage{lineno}
\usepackage{natbib}
\usepackage{pdflscape}
\usepackage{hyperref}
\usepackage{multirow}

\usepackage{titlesec}
\titlespacing*{\subsubsection}{0pt}{0.5ex}{0pt}
\titlespacing*{\subsection}{0pt}{1.0ex}{0pt}
\usepackage{changepage}

\makeatother

\begin{document}
\begin{flushleft}
\setcounter{footnote}{0}
\begin{spacing}{1.9}
\begin{flushleft}Running Head: Autocorrelation in ecological data
\par\end{flushleft}

\noindent \textbf{The basis function approach for modeling autocorrelation
in ecological data}

\medskip{}

\noindent \textbf{Trevor J. Hefley}\footnote{Corresponding author; E-mail: Trevor.Hefley@colostate.edu.}

\noindent {\small{}Department of Fish, Wildlife, and Conservation
Biology, Colorado State University}{\small \par}

\noindent {\small{}Department of Statistics, Colorado State University}{\small \par}

\medskip{}

\noindent \textbf{Kristin M. Broms}

\noindent {\small{}Department of Fish, Wildlife, and Conservation
Biology, Colorado State University}{\small \par}

\medskip{}

\noindent \textbf{Brian M. Brost}

\noindent {\small{}Department of Fish, Wildlife, and Conservation
Biology, Colorado State University}{\small \par}

\medskip{}
\textbf{Frances E. Buderman}

\noindent {\small{}Department of Fish, Wildlife, and Conservation
Biology, Colorado State University}{\small \par}

\medskip{}
\textbf{Shannon L. Kay}

\noindent {\small{}Department of Statistics, Colorado State University}{\small \par}

\medskip{}
\textbf{Henry R. Scharf}

\noindent {\small{}Department of Statistics, Colorado State University}{\small \par}

\medskip{}
\textbf{John R. Tipton}

\noindent {\small{}Department of Statistics, Colorado State University}{\small \par}

\medskip{}
\textbf{Perry J. Williams}

\noindent {\small{}Department of Fish, Wildlife, and Conservation
Biology, Colorado State University}{\small \par}

\noindent {\small{}Department of Statistics, Colorado State University}{\small \par}

\medskip{}

\noindent \textbf{Mevin B. Hooten}

\noindent {\small{}U.S. Geological Survey, Colorado Cooperative Fish
and Wildlife Research Unit}{\small \par}

\noindent {\small{}Department of Fish, Wildlife, and Conservation
Biology, Colorado State University}{\small \par}

\noindent {\small{}Department of Statistics, Colorado State University}{\small \par}

\bigskip{}

\setlength{\parindent}{0.7cm}

\begin{doublespace}

\section*{\textmd{\textsc{\normalsize{}Abstract}}}
\end{doublespace}

\noindent Analyzing ecological data often requires modeling the autocorrelation
created by spatial and temporal processes. Many of the statistical
methods used to account for autocorrelation can be viewed as regression
models that include basis functions. Understanding the concept of
basis functions enables ecologists to modify commonly used ecological
models to account for autocorrelation, which can improve inference
and predictive accuracy. Understanding the properties of basis functions
is essential for evaluating the fit of spatial or time-series models,
detecting a hidden form of multicollinearity, and analyzing large
data sets. We present important concepts and properties related to
basis functions and illustrate several tools and techniques ecologists
can use when modeling autocorrelation in ecological data. \bigskip{}

\noindent \emph{Key words: autocorrelation; Bayesian model; collinearity;
dimension reduction; semiparametric regression; spatial statistics;
time series}

\subsection*{\textmd{\textsc{\normalsize{}Introduction}}}

Ecological processes interact at multiple temporal and spatial scales,
generating complex spatio-temporal patterns (\citealt{levin1992problem}).
The science of ecology is concerned with understanding, describing,
and predicting components of these spatio-temporal patterns using
limited and noisy observations. An important consideration when developing
an ecological model is how to include the spatial, temporal, or spatio-temporal
aspects of the process (\citealt{legendre1993spatial}). For example,
species distribution models are used to understand and predict how
the abundance of plants and animals varies across space and time (\citealt{elith2009species}).
The abundance of a species within a patch of habitat might depend
on environmental covariates (e.g., minimum annual temperature), but
might also depend on the abundance in surrounding patches. Dispersal
of organisms among habitat patches can make abundance in nearby patches
more similar than could be explained by environmental covariates alone
(\citealt{legendre1989spatial}). Correlation among observations that
depends on the proximity (in space, time, or both) of neighboring
observations is known as autocorrelation (Table 1). Disentangling
autocorrelation from environmental covariates is critical to understanding
endogenous and exogenous factors that influence populations and ecosystems
(\citealt{borcard1992partialling}). Moreover, properly accounting
for autocorrelation is necessary for obtaining reliable statistical
inference (e.g., \citealt{Fieberg2012}).

Isolating the effect of autocorrelation in an ecological model can
be accomplished by including a function that captures the dependence
between observations that are close in space or time. The mathematical
form of the function that best describes the autocorrelation is always
unknown and may be complex, but can be approximated by a combination
of simple basis functions. Most ecologists have encountered basis
functions (e.g., polynomial regression), but may not be aware of the
breadth of situations in which they can be used to model autocorrelation
in ecological data. For example, basis functions are used in semiparametric
models, such as generalized additive models, but are also implicitly
used in spatial or time-series models. Understanding how basis functions
can be used to model autocorrelation is essential for evaluating the
fit of spatial, time-series, or spatio-temporal models, detecting
a hidden form of multicollinearity (\citealt{Hodges2010}), and facilitating
the analysis of large data sets (\citealt{wikledimred}). More importantly,
understanding the basis function approach will enable ecologists to
tailor many commonly used models to account for autocorrelation, which
can improve both inference and predictive accuracy (e.g., \citealt{hooten2003predicting};
\citealt{conn2015using,buderman2015functional}).

We have three goals in this paper: 1) introduce concepts and terminology
related to basis functions and autocorrelation; 2) demonstrate the
connections between commonly used methods to model autocorrelation;
and 3) develop a working knowledge of the basis function approach
so ecologists can devise ways to model autocorrelation in commonly
used ecological models. We first introduce basis functions and then
the concepts of first-order and second-order model specifications.
To illustrate these concepts, we present three examples: a standard
regression model, a time-series model, and a spatial model, each applied
to different types of ecological data. We include supplementary material
comprised of a tutorial that contains additional descriptions, data,
and example computer code.

\subsection*{\textmd{\textsc{\normalsize{}Basis Functions}}}

Consider a simple linear regression model of pelagic bioluminescence
density (sources) as a function of water depth (Fig. 1a; \citealt{gillibrand2007seasonal})
\begin{equation}
\mathbf{y}=\alpha_{0}\mathbf{z}_{0}+\alpha_{1}\mathbf{z}_{1}+\boldsymbol{\mathbf{\boldsymbol{\varepsilon}}}\:,\label{eq:poly1-1}
\end{equation}

\noindent where $\mathbf{y}$ is an $n\times1$ vector of bioluminescence
density, $\mathbf{z}_{0}$ is an $n\times1$ vector of ones, $\mathbf{z}_{1}$
is an $n\times1$ vector that contains the depth in meters of the
observed bioluminescence sources, $\alpha_{0}$ is the intercept,
$\alpha_{1}$ is a regression coefficient, and $\boldsymbol{\mathbf{\boldsymbol{\varepsilon}}}$
is an $n\times1$ vector that contains independent and normally distributed
error terms with variance $\sigma_{\varepsilon}^{2}$ (i.e., $\boldsymbol{\mathbf{\boldsymbol{\varepsilon}}}\sim\text{N}(\mathbf{0}\textnormal{, }$$\sigma_{\varepsilon}^{2}\mathbf{I}$),
where $\mathbf{I}$ is the $n\times n$ identity matrix). In this
simple linear regression model, the basis coefficients are $\alpha_{0}$
and $\alpha_{1}$ and the basis vectors $\mathbf{z}_{0}$ and $\mathbf{z}_{1}$
are the depths raised to the power $0$ and 1 (Table 1). It may not
be common to refer to transformations of a covariate as basis vectors;
however, the transformations form a ``basis'' of possible values
in covariate space. The function that transforms a covariate into
a basis vector is referred to as a basis function. Although the terms
basis function and basis vector tend to be used interchangeably in
the literature, basis functions typically exist in continuous space,
whereas basis vectors are the output from a function and exist in
discrete space (e.g., where depth was measured; Table 1; see Appendix
A for further discussion). All transformations of a covariate are
known collectively as a basis expansion (e.g., $\mathbf{z}_{0}$ and
$\mathbf{z}_{1}$). Finally, as in simple linear regression, the expected
density of bioluminescence at the measured depths is the linear combination
of basis vectors $\alpha_{0}\mathbf{z}_{0}+\alpha_{1}\mathbf{z}_{1}$
(Fig. 1a).

It is clear from Fig. 1a that the simple linear regression model does
not adequately capture the relationship between bioluminescence and
depth. A more flexible basis expansion that better captures the relationship
is the polynomial regression model that includes the quadratic effect
of depth 
\begin{equation}
\boldsymbol{\mathbf{y}}=\alpha_{0}\mathbf{z}_{0}+\alpha_{1}\mathbf{z}_{1}+\alpha_{2}\mathbf{z}_{2}+\boldsymbol{\mathbf{\boldsymbol{\varepsilon}}}\:,\label{eq:poly2-1}
\end{equation}
where $\alpha_{2}$ is the basis coefficient for the squared effect
of depth ($\mathbf{z}_{2}$; Fig. 1b). Some models that use basis
functions can be respecified, which can aid in interpretation and
increase computational efficiency and stability. For example, we can
respecify Eq. 2 using a different basis expansion, but in a way that
yields the exact same model
\begin{equation}
\mbox{\ensuremath{\mathbf{y}}}=\alpha_{1}(\mathbf{z}_{1}-k_{1})^{2}+\alpha_{2}(\mathbf{z}_{1}-k_{2})^{2}+\alpha_{3}(\mathbf{z}_{1}-k_{3})^{2}+\boldsymbol{\mathbf{\boldsymbol{\varepsilon}}}\:,\label{eq:poly3-1}
\end{equation}
where $\mathbf{z}_{1}$ is an $n\times1$ vector of the observed depths
in meters, $k_{j}$ is the $j^{th}$ depth of interest ($j=1,2,3)$,
and $\alpha_{j}$ is the basis coefficient. The two basis expansions
in Eq. \ref{eq:poly2-1} and Eq. \ref{eq:poly3-1} have different
basis vectors and will yield different estimates of $\alpha_{j}$,
but result in the exact same polynomial curve when fit to the data.
For example, let $k_{1}=1140$ m, $k_{2}=2620$ m, and $k_{3}=3420$
m, and compare the basis vectors and predicted bioluminescence (cf.,
Fig. 1b and 1c). An interactive figure of this example can be found
in Appendix B.

Even if the specifications result in the exact same model, there are
many reasons why one basis expansion might be preferred over others.
For example, the number of parameters in the model can be reduced
if a basis expansion results in some basis coefficients that can be
assumed to be zero (see confidence intervals for $\alpha_{j}$ in
Appendix B). In addition, some basis expansions might have scientific
interpretations. For example, the model in Eq. \ref{eq:poly3-1} states
that the density of bioluminescence is a function of the distance
between the observed depth and three locations in the water column,
that we might believe are biologically important. Finally, some basis
expansions may reduce the correlation among basis vectors. For example,
the coefficient of determination ($R^{2}$) for the basis vectors
$\mathbf{z}_{1}$ and $\mathbf{z}_{2}$ in Eq. 2 is 0.96, whereas,
the maximum $R^{2}$ among the three basis vectors in Eq. 3 is 0.25.

\subsubsection*{\textmd{\textit{Model assumptions}}}

A critical assumption is that a linear combination of basis vectors
adequately approximates the unknown relationship between the observed
response and basis vectors. In a regression context, this assumption
is analogous to assuming the model includes all covariates that linearly
influence the response. For example, it is assumed in Eq. \ref{eq:poly2-1}
and Eq. \ref{eq:poly3-1} that a linear combination of the basis vectors
(e.g., $\alpha_{0}\mathbf{z}_{0}+\alpha_{1}\mathbf{\mathbf{z}}_{1}+\alpha_{2}\mathbf{z}_{2}$)
adequately represents the unknown functional relationship between
depth and the density of bioluminescence. A result of this assumption
is that the basis vectors must span the space of interest, otherwise
the model is inadequate. In the bioluminescence example, the space
of interest is depth and the bases in Eq. 2 and Eq. 3 span the set
of all second-degree polynomial functions. That is, we assume the
true underlying relationship between the density of bioluminescence
and depth can be appropriately modeled as a second-degree polynomial.
Because Eq. \ref{eq:poly2-1} and Eq. \ref{eq:poly3-1} are both second-degree
polynomials and the two basis expansions span the same space, the
estimated curves are exactly the same (cf., Fig. 1b and 1c). If one
or more of the basis vectors were removed from the polynomial regression
model, the basis vectors would not span the same space of all second-degree
polynomials and therefore could not create the exact curve shown in
Fig. 1b and 1c.

\subsubsection*{\textmd{\textit{Generalizations}}}

Now consider an unknown function $\eta(x)$ that describes a pattern
or process in nature that generates autocorrelation. This function
can be continuous or discrete, but the true form is always unknown.
For example, $\eta(x)$ might describe the similarity in abundance
between habitat patches in geographic space, population regulation
influenced by endogenous factors in temporal space (e.g., density
dependence), or how net primary productivity changes with temperature
in covariate space. Although the true form of the function $\eta(x)$
is unknown, we can approximate it with a combination of simple basis
functions. We can combine the basis functions in a linear equation
such that $\eta(x)\approx\sum_{j=1}^{m}\alpha_{j}f_{j}(x)$; this
is a general notation that contains $m$ basis functions $f_{j}(x)$
($j=1,...,m)$. For example, in the polynomial regression model (Eq.
\ref{eq:poly2-1}), $f_{1}(x)=x^{0}$, $f_{2}(x)=x^{1}$, and $f_{3}(x)=x^{2}$.
In what follows, we use matrix notation and write $\boldsymbol{\eta}\equiv\mathbf{Z}\boldsymbol{\alpha}$,
where $\boldsymbol{\eta}$ is an $n\times1$ vector representing an
approximation of the unknown function $\eta(\mathbf{x})$ at the $n$
locations $\mbox{\ensuremath{\mathbf{x}}}$ in the space of interest,
$\mathbf{Z}$ is an $n\times m$ matrix containing the basis vectors,
and $\boldsymbol{\alpha}$ is an $m\times1$ vector of basis coefficients.
We also use the matrix notation $\mathbf{X}\boldsymbol{\beta}$, where
$\mathbf{X}$ is an $n\times p$ matrix of traditional covariates
and $\boldsymbol{\beta}$ is a $p\times1$ vector of traditional regression
coefficients. Our notation is used to identify when a basis vector
should be treated as a traditional covariate (and included in $\mathbf{X}$)
or treated as a formal basis expansion (and included in $\mathbf{Z}$).
In some applications, there is no practical difference between including
basis vectors in $\mathbf{X}$ or $\mathbf{Z}$, whereas, in other
applications, the choice of notation is used to designate whether
the coefficients are treated as fixed ($\mathbf{X}$) or random ($\mathbf{Z}$)
effects.

\subsection*{\textmd{\textsc{\normalsize{}Model Specifications}}}

Specifying a statistical model involves combining probability distributions
and deterministic mathematical equations, both with unknown parameters,
in a way that properly describes the ecological process (\citealt{Hobbs2015}).
As we demonstrated in the polynomial regression example, there are
multiple ways to specify the exact same model (e.g., Eq. 2 and Eq.
3). An important concept related to specifying a model is first-order
and second-order model specifications (Table 1; \citealt{cressie2011statistics,Hobbs2015}),
which is also known in the mixed-model literature as G-side and R-side
specifications (\citealt{littell2006sas,Stroup2012}). Specifically,
this concept relates to placing a function that describes the autocorrelation
in either the mean (expected value) or the covariance of a probability
distribution. The concept of first-order and second-order model specification
is important for understanding the equivalence of certain spatial
and time-series models, efficient implementation of Bayesian models
that account for autocorrelation, and for detecting a hidden form
of multicollinearity. A general understanding of hierarchical models
or mixed models is necessary for what follows. Introductions to hierarchical
models or mixed models include: \citet{Hobbs2015}, \citet{hodges2013richly},
\citet{littell2006sas}, \citet{Ruppert2003}, \citet{Stroup2012},
\citet{Wood2006}, and \citet{Zuur2009}.

\subsubsection*{\textmd{\textit{First-order specification}}}

A first-order specification defines a function that describes the
autocorrelation in the mean structure of a distribution (Table 1).
For example, consider the linear regression model $\mathbf{y}=\textnormal{\ensuremath{\boldsymbol{X\beta}}}+\boldsymbol{\varepsilon}$,
where $\mathbf{y}\equiv(y_{1},\ldots,y_{n})'$ are the observed data. Assuming
independent and normally distributed errors ($\boldsymbol{\varepsilon}\sim\text{N}(\mathbf{0},\sigma_{\varepsilon}^{2}\mathbf{I})$),
we can write this model as 
\begin{equation}
\mathbf{y}\sim\text{N}(\mbox{\ensuremath{\mathbf{{X}\mathbf{\boldsymbol{{\beta}}}\textnormal{, }}\sigma_{\varepsilon}^{2}\mathbf{I}}})\:.\label{eq:first-order2-1}
\end{equation}
An assumption of Eq. 4 is that a linear combination of the covariates serves
as a good approximation to the underlying relationship. If there is
evidence that the model in Eq. 4 is inadequate (e.g., correlation
among the residual errors), a linear combination of basis vectors
may be added to the mean to improve model fit and satisfy model assumptions,
such that 
\begin{equation}
\mbox{\ensuremath{\mathbf{y}}}\sim\text{N}(\mathbf{{X}\boldsymbol{{\beta}}+{Z}\mathbf{\boldsymbol{{\alpha}}\textnormal{, }}\sigma_{\varepsilon}^{\textnormal{2}}\mathbf{I}})\:.\label{eq:first-order3-1}
\end{equation}
 The basis expansion ($\mathbf{Z}$) account for additional complexity
in the mean structure that is not explained by the covariates in $\mathbf{X}$.
For example, $\mathbf{Z}$ could account for the lack of fit (spatial
autocorrelation) in the linear or polynomial model in the bioluminescence
example (Fig. 1; Table 1).

\subsubsection*{\textmd{\textit{Second-order specification}}}

A second-order specification defines a function that describes the
autocorrelation in the covariance of a probability distribution. For
example, consider the linear model $\mathbf{y}=\textnormal{\ensuremath{\boldsymbol{X\beta}}}+\mathbf{\boldsymbol{\eta}}+\boldsymbol{\varepsilon}$
where $\boldsymbol{\varepsilon}\sim\text{N}(\mathbf{0},\sigma_{\varepsilon}^{2}\mathbf{I})$,
$\boldsymbol{\eta}\sim\text{N}(\mathbf{0},\sigma_{\alpha}^{2}\mathbf{R})$,
and the random effect $\mathbf{\boldsymbol{\eta}}$ results in correlated
errors. We can write the model as:
\begin{equation}
\mbox{\ensuremath{\mathbf{y}}}\sim\text{N}(\mathbf{X}\boldsymbol{\beta}\textnormal{, }\sigma_{\varepsilon}^{2}\mathbf{I}+\sigma_{\alpha}^{2}\mbox{\ensuremath{\mathbf{R}}})\:,\label{eq:second-order1-1}
\end{equation}

\noindent where $\mathbf{R}$ is a correlation matrix that accounts
for autocorrelation among observations. The correlation matrix $\mathbf{R}$
is often specified using a correlation function that depends on a
distance measure between two observations in the space of interest
(Table 1).

\subsubsection*{\textmd{\textit{Equivalent specifications}}}

In some situations, the first-order and second-order specifications
result in the same model. To demonstrate this for a specific case,
we make the additional assumption that the basis coefficients in Eq.
5 are normally-distributed random effects (i.e., $\boldsymbol{\alpha}\sim\text{N}(\mathbf{0},\sigma_{\alpha}^{2}\mathbf{I})$).
Equivalent probability density functions can be obtained by integrating
out the random effect $\boldsymbol{\alpha}$ from the first-order
specification in Eq. \ref{eq:first-order3-1} to yield 
\begin{equation}
\begin{aligned}\mathbf{y} & \sim\int\text{N}(\mathbf{X}\boldsymbol{\beta}+\mathbf{Z\mathbf{\boldsymbol{\alpha}}},\sigma_{\varepsilon}^{2}\mathbf{I})\text{N}(\mathbf{0},\sigma_{\alpha}^{2}\mathbf{I})d\boldsymbol{\alpha}\\
 & =\text{N}(\mathbf{X}\boldsymbol{\beta},\sigma_{\varepsilon}^{2}\mathbf{I}+\sigma_{\alpha}^{2}\mbox{\ensuremath{\mathbf{\mathbf{Z}\mathbf{Z}}'}})\:,
\end{aligned}
\label{eq:second-order2-1-1}
\end{equation}
where equivalence between the first- and second-order specifications
holds if the correlation matrix $\mathbf{R}$ is the outer product
of the basis expansion $\mathbf{Z}$ (i.e., $\mathbf{R}\equiv\mathbf{Z}\mathbf{Z}'$;
see example below). The integration in Eq. 7 effectively ``moves''
the autocorrelation modeled by the basis vectors in the mean structure
to the covariance structure. For example, consider a mixed-effects
model where $\mathbf{Z}$ is used to represent autocorrelation due
to a site or grouping effect among observations: 

\begin{equation}
\mathbf{Z}=\left[\begin{array}{ccc}
1 & 0 & 0\\
1 & 0 & 0\\
0 & 1 & 0\\
0 & 1 & 0\\
0 & 0 & 1\\
0 & 0 & 1
\end{array}\right]\:,\label{eq:z cmpd sym-1}
\end{equation}

\noindent where, $y_{1}$ and $y_{2}$ were observed at the first
site, $y_{3}$ and $y_{4}$ were observed at the second site, etc.
If we assume the basis coefficients are normally-distributed random
effects, then 
\begin{equation}
\mathbf{R}=\mathbf{Z}\mathbf{Z}'=\left[\begin{array}{ccc}
1 & 0 & 0\\
1 & 0 & 0\\
0 & 1 & 0\\
0 & 1 & 0\\
0 & 0 & 1\\
0 & 0 & 1
\end{array}\right]\left[\begin{array}{cccccc}
1 & 1 & 0 & 0 & 0 & 0\\
0 & 0 & 1 & 1 & 0 & 0\\
0 & 0 & 0 & 0 & 1 & 1
\end{array}\right]=\left[\begin{array}{cccccc}
1 & 1 & 0 & 0 & 0 & 0\\
1 & 1 & 0 & 0 & 0 & 0\\
0 & 0 & 1 & 1 & 0 & 0\\
0 & 0 & 1 & 1 & 0 & 0\\
0 & 0 & 0 & 0 & 1 & 1\\
0 & 0 & 0 & 0 & 1 & 1
\end{array}\right]\:\:,\label{eq:z cmpd sym-1}
\end{equation}

\noindent where $\mathbf{R}$ is called the compound symmetry correlation
matrix (\citealt{littell2006sas,Zuur2009,Stroup2012}). The model
that is obtained by using the first-order specification in Eq. 8 is
the exact same model that would be obtained by using the correlation
matrix in Eq. 9.

Although one may start with a basis expansion $\mathbf{Z}$, many
methods developed to model autocorrelation start by choosing a correlation
matrix $\mathbf{R}$. When starting with a correlation matrix $\mathbf{R}$,
a basis expansion $\mathbf{Z}$ can be obtained using matrix decomposition
methods (e.g., spectral decomposition; \citealt{Lorenz1956}; \citealt{cressie2011statistics};
pg. 156). For example, consider the regression model in Eq. 6, and
let $\mathbf{R}(\phi)$ be an order-one autoregressive correlation
matrix (AR(1))
\begin{equation}
\mathbf{R(\phi})=\left[\begin{array}{ccccc}
1 & \phi^{1} & \phi^{2} & \cdots & \phi^{t-1}\\
\phi^{1} & 1 & \phi^{1} & \cdots & \phi^{t-2}\\
\phi^{2} & \phi^{1} & 1 & \vdots & \phi^{t-3}\\
\vdots & \vdots & \vdots & \ddots & \ddots\\
\phi^{t-1} & \phi^{t-2} & \phi^{t-3} & \ddots & 1
\end{array}\right]\:,\label{eq:timeseries2-1}
\end{equation}
where $-1<\phi<1$. The AR(1) correlation matrix in Eq. 10 is commonly
used in time-series analysis to model temporal correlation that diminishes
exponentially with a rate of decay that depends on $\phi$. When a
correlation matrix, or basis expansion depends on parameters, we include
the parameters in parentheses (e.g., $\mathbf{R}(\phi)$ and $\mathbf{Z}(\phi)$). 

A correlation matrix can be decomposed to produce basis vectors that
are useful in the first-order specification. One approach to obtain
basis vectors from $\mathbf{R}(\phi)$ is the spectral decomposition:
$\mathbf{\mbox{\ensuremath{\mathbf{R}}(\ensuremath{\phi})}}=\mathbf{Q\Lambda}\mathbf{Q}'$,
where $\mathbf{Q}$ are the eigenvectors and $\boldsymbol{\Lambda}$
is a diagonal matrix with elements that contain the eigenvalue associated
with each eigenvector (\citealt{cressie2011statistics}; pgs. 156-157).
Using the spectral decomposition, the basis expansion can be written
as $\mathbf{Z}(\phi)=\mathbf{Q\Lambda}^{\frac{1}{2}}$. As an example,
for three observations $\mathbf{y}\equiv(y_{1},y_{2},y_{3})'$ collected
at times $t=1,2,3$, the AR(1) correlation matrix using $\phi=0.5$
is

\begin{equation}
\mathbf{R}(\phi)=\left[\begin{array}{ccc}
1 & 0.5 & 0.25\\
0.5 & 1 & 0.5\\
0.25 & 0.5 & 1
\end{array}\right]\:.\label{eq:z cmpd sym}
\end{equation}

\noindent The spectral decomposition of $\mathbf{R}(\phi)$ is

\begin{equation}
\mathbf{\mbox{\ensuremath{\mathbf{R}}(\ensuremath{\phi})}}=\mathbf{Q\Lambda}\mathbf{Q}'=\left[\begin{array}{ccc}
-0.54 & -0.71 & 0.45\\
-0.64 & 0 & -0.77\\
-0.54 & 0.71 & 0.45
\end{array}\right]\left[\begin{array}{ccc}
1.84 & 0 & 0\\
0 & 0.75 & 0\\
0 & 0 & 0.41
\end{array}\right]\left[\begin{array}{ccc}
-0.54 & -0.64 & -0.54\\
-0.71 & 0 & 0.71\\
0.45 & -0.77 & 0.45
\end{array}\right]\:.\label{eq:z cmpd sym-2}
\end{equation}

\noindent The matrices of eigenvectors $(\mbox{\ensuremath{\mathbf{Q}}}$)
and eigenvalues ($\boldsymbol{\Lambda}$) in Eq. 12 can be used to
construct the basis expansion

\begin{equation}
\mathbf{Z}(\phi)=\mathbf{Q\Lambda}^{\frac{1}{2}}=\left[\begin{array}{ccc}
-0.74 & -0.61 & 0.29\\
-0.87 & 0 & -0.49\\
-0.74 & 0.61 & 0.29
\end{array}\right]\:.\label{eq:z cmpd sym-2-1}
\end{equation}

\noindent Alternatively, one might use the eigenvectors $\mathbf{Q}$
as basis vectors (i.e., $\mathbf{Z}(\phi)\equiv\mathbf{Q}(\phi)$;
\citealt{griffith2006spatial}), which would require specifying a
non-constant variance for the basis coefficients such that $\boldsymbol{\alpha}\sim\text{N}(\mathbf{0},\sigma_{\alpha}^{2}\mathbf{\boldsymbol{\Lambda}})$.
Converting models between first- and second-order specifications using
the techniques we have presented is critical for harnessing the full
power of basis functions for modeling autocorrelation in ecological
data.

\subsubsection*{\textmd{\textit{Generalized models}}}

Basis functions can be used with any response distribution (e.g.,
Poisson, binomial). For example, the generalized linear mixed model
can include random effects for the coefficients associated with the
basis vectors that account for autocorrelation (\citealt{Bolker2009}).
Similarly, basis functions can also be embedded within Bayesian hierarchical
models to account for autocorrelation (see Example 3). In all of the
examples presented below, we assume that the basis coefficients are
normally-distributed random effects; however, this is not a necessary
assumption. As with the generalized linear mixed model or Bayesian
hierarchical models, distributions for random effects other than the
normal could be used (e.g., gamma, t-distribution; \citealt{Higdon2002,lee2006generalized,johnson2013bayesian,Gelman2013,Hobbs2015}).

\subsection*{\textmd{\textsc{\normalsize{}Example 1: pelagic bioluminescence versus
depth gradient}}}

In the Basis Functions section, we initially modeled the density of
bioluminescence using depth or transformations of depth as covariates
with coefficients that were fixed effects. Depth can also be thought
of as the spatial location in the water column, so it is natural to
model the density of bioluminescence using a spatial model instead
of regressing on depth directly. In the three following model specifications
that we use to capture spatial autocorrelation, $\mathbf{X}$ is an
$n\times1$ matrix of ones and $\mathbf{\boldsymbol{\beta}}$ is a
constant intercept term (i.e., $\mathbf{X}\boldsymbol{\beta}=\beta_{0}$).
As a result, the influence of depth will be modeled in either the
basis expansion $\mathbf{Z}$ or the correlation matrix $\mathbf{R}$,
in accordance with the first-order or second-order specification,
respectively.

\subsubsection*{\textmd{\textit{Spatial regression model: a second-order specification}}}

Consider the model in Eq. \ref{eq:second-order1-1} where the correlation
matrix $\mathbf{R}(\phi)$ is specified using a parametric correlation
function that depends on a range parameter $\phi$ (\citealt{cressie2011statistics,banerjee2014hierarchical}).
The range parameter controls how correlation diminishes as the distance
between locations increases. For this example, we use a Gaussian correlation
function 
\begin{equation}
r_{ij}(\phi)=e^{-\frac{d_{ij}^{2}}{\phi}}\:,\label{eq:spatial2}
\end{equation}

\noindent where $d_{ij}$ is the distance between locations $i$ and
$j$ (note that $d_{ij}=0$ for $i=j$) and $r_{ij}(\phi)$ is the
element in the $i^{th}$ row and $j^{th}$ column of $\mathbf{R}(\phi)$.
For the bioluminescence example, $d_{ij}$ is the difference in depth
between observations $i$ and $j$. Given the second-order specification,
it is not immediately clear how to estimate the influence of depth
on bioluminescence (i.e., $\beta_{0}+\boldsymbol{\eta}$), which requires
the spatial random effect $\boldsymbol{\eta}$. To predict bioluminescence
at the observed (and unobserved) depths using the second-order specification,
we used best linear unbiased prediction (BLUP; Kriging in the spatial
context; see \citealt{robinson1991blup} for derivation). The predicted
spatial random effect ($\boldsymbol{\eta}$) for the observed depths,
given estimates of all other parameters, can be obtained using
\begin{equation}
\mathbf{\hat{\boldsymbol{\eta}}}=\mathbf{\hat{\sigma}}_{\alpha}^{2}\mathbf{R}(\hat{\phi})\left(\hat{\sigma}_{\varepsilon}^{2}\mathbf{I}+\hat{\sigma}_{\alpha}^{2}\mbox{\ensuremath{\mathbf{R}}(\ensuremath{\hat{\phi}})}\right)^{-1}\left(\mbox{\ensuremath{\mathbf{y}-}}\mathbf{X}\hat{\boldsymbol{\beta}}\right)\:.\label{eq:spatial3}
\end{equation}

\noindent The fitted spatial model (Fig. 2a; $\hat{\beta}_{0}+\hat{\boldsymbol{\eta}}$)
captures fine scale (local) variability better than the polynomial
regression model (Fig. 1b and 1c; Appendix C).

\subsubsection*{\textmd{\textit{Spatial regression model: a first-order specification}}}

The spatial regression model can also be implemented using basis vectors.
Consider the first-order specification from Eq. \ref{eq:first-order3-1}
where $\boldsymbol{\alpha}\sim\text{N}(\mathbf{0},\sigma_{\alpha}^{2}\mathbf{I})$
and $\mathbf{Z}(\phi)$ is obtained from a spectral decomposition
of $\mathbf{R}(\phi)$. Using the BLUP from Eq. 15, basis coefficients
are equivalent to 
\begin{equation}
\hat{\boldsymbol{\alpha}}=\mathbf{\hat{\sigma}}_{\alpha}^{2}\mathbf{Z}(\hat{\phi})^{'}\left(\hat{\sigma}_{\varepsilon}^{2}\mathbf{I}+\hat{\sigma}_{\alpha}^{2}\mbox{\ensuremath{\mathbf{Z}}(\ensuremath{\hat{\phi}})\ensuremath{\mathbf{Z(\hat{\phi})}}}^{'}\right)^{-1}\left(\mbox{\ensuremath{\mathbf{y}-}}\mathbf{X}\hat{\boldsymbol{\beta}}\right)\:.\label{eq:spatial6}
\end{equation}
Note that because $\mathbf{R}(\hat{\phi})\equiv\mathbf{Z}(\hat{\phi})\mathbf{Z(\mathbf{\hat{\phi}})}'$,
by definition $\mathbf{Z}(\hat{\phi})\hat{\mathbf{\boldsymbol{\alpha}}}$
is the same as $\hat{\boldsymbol{\eta}}$ in Eq. \ref{eq:spatial3}.
The bioluminescence for any given depth (i.e., $\mathbf{\hat{\beta}_{\text{0}}}+\mathbf{Z}(\hat{\phi})\hat{\boldsymbol{\alpha}})$
from the first-order specification is shown in Fig. 2b along with
three of the eigen basis vectors multiplied by the corresponding basis
coefficients. The fitted values from the first- and second-order specifications
are exactly the same (cf., Fig. 2a to 2b) because both specifications
result in an equivalent model. 

Even if the initial model formulation is a second-order specification
that uses a correlation function, the equivalent first-order specification
is often useful. Three important uses for the first-order specification
are: 1) it allows for assessment of collinearity between covariates
in $\mathbf{X}$ and basis vectors in $\mathbf{Z}(\phi)\:$(see Example
2); 2) basis vectors can be visualized and certain types of basis
expansions have useful ecological interpretation (e.g., \citealt{griffith2006spatial});
and 3) certain types of basis expansions are useful for dimension
reduction required to fit models to large data sets \citep{wikledimred}.
We will demonstrate the utility of the first-order specification in
the following examples.

\subsubsection*{\textmd{\textit{Modeling spatial autocorrelation using kernel basis
functions}}}

Another method that can be used to model autocorrelation is kernel
regression. Kernel regression is a semiparametric regression technique
widely used by statisticians and the machine learning community \citep{bishop2006pattern,Hastie2009,James2013}.
Regression models that employ kernel basis functions are written using
the first-order specification. The commonly used Gaussian kernel basis
function is defined as: 
\begin{equation}
z_{ij}(\phi)=e^{-\frac{2d_{ij}^{2}}{\phi}}\:,\label{eq:kernel1}
\end{equation}

\noindent where $z_{ij}(\phi)$ is the element in the $i^{th}$ row
and $j^{th}$ column of $\mathbf{Z}(\phi)$, and $d_{ij}$ is the
distance between the $i^{th}$ data point and the $j^{th}$ knot ($j=1,\dots,m$
where $m$ is the number of basis vectors). Knots are locations in
the space of interest where each basis vector is anchored (e.g., $k_{j}$
in the bioluminescence example; Table 1). In Fig. 2c, we show the
density of bioluminescence predicted from the kernel regression model
(i.e., $\mathbf{\hat{\beta}}_{0}+\mathbf{Z}(\hat{\phi})\hat{\boldsymbol{\alpha}}$)
and three of its basis vectors (from a total $m=17)$ multiplied by
the corresponding basis coefficients estimated using Eq. \ref{eq:spatial6}.
Comparison of the eigen and kernel basis vectors reveals that the
two types of basis functions look different, but the fitted curves
are nearly equivalent (cf., Fig. 2b and 2c). Importantly, as the number
of basis vectors and knots increases to infinity (on a grid), the
first-order model specification that uses a Gaussian kernel basis
function (Eq. 17) converges to the second-order specification that
uses a Gaussian correlation function (Eq. 14; \citealt{Higdon2002}).
An interactive figure that allows users to experiment with basis functions
for this example can be found in Appendix B. 

Regression models that use kernel basis functions are useful because
they allow for more complex correlation structures compared to models
that rely on correlation functions (\citealt{Higdon2002,sampsonnonstationarykernel};
Table 2). Further, the number of basis vectors and coefficients can
be controlled by the user depending on the level of computational
efficiency required. Choosing the dimension of the basis expansion
($m$) to be less than $n$ is known as dimension reduction. For example,
the spatial model that relies on a second-order specification and
uses a correlation function can be converted to a first-order model
that requires 51 eigen basis vectors, but the kernel regression uses
a pre-selected number of kernel basis vectors (17 for this example).
Dimension reduction usually relies on first-order model specification
and is helpful for modeling autocorrelation in large data sets, allowing
for statistical inference that would otherwise be computationally
infeasible.

\subsection*{\textmd{\textsc{\normalsize{}Example 2: Population Trend }}}

A common task in applied ecology is to infer if the abundance of a
species is declining over time. This task often involves fitting a
trend to a time series of abundance indices and inferring if the associated
regression coefficient is negative. For this example, we use the simple
linear model
\begin{equation}
\mbox{\ensuremath{\mathbf{y}}}\sim\text{N}(\beta_{0}+\beta_{1}\mathbf{t},\sigma_{\varepsilon}^{2}\mathbf{I})\:,\label{eq:timeseries1-1}
\end{equation}
where $\mathbf{y}$ is a $t\times1$ vector containing the count or
index of population size from each time period, and $\mathbf{t}$
is the corresponding vector of times.

Bobwhite quail (\textit{Colinus virginianus}) are a common species
that occurs throughout a large portion of the United States, but are
declining in abundance in many regions (\citealt{veech2006increasing}).
We present an index of bobwhite quail population size obtained from
annual whistle-count surveys in Nemaha County, Nebraska (Fig. 3a;
see \citealt{hefley2013statistical} for a detailed description of
the data). We fit a simple linear regression model to these data using
maximum likelihood which results in $\hat{\beta}_{1}=-1.16$ and a
95\% confidence interval of $[\textnormal{-1.88, -0.44]}$. The fitted
regression model suggests a decline in abundance; however, a clear
pattern in the residuals is present, possibly due to underlying population
dynamics of bobwhite quail (Fig. D1; Appendix D). If autocorrelation
is ignored, the uncertainty associated with regression coefficients
will be underestimated and may cause the researcher to overstate the
statistical significance of the decline (\citealt{cressie1993statistics,Hoeting2009}).
The underlying population dynamics that generated the autocorrelation
are likely complex and building mechanistic models that account for
the process is challenging (\citealt{hefley2013statistical}). A simpler
approach for modeling the endogenous population dynamics is to assume
that the population size (or some transformation thereof) can be modeled
as $\boldsymbol{\mathbf{y}}\sim\text{N}(\beta_{0}+\beta_{1}\mathbf{t},\sigma_{\alpha}^{2}\mathbf{R(\phi})+\sigma_{\varepsilon}^{2}\mathbf{I}$),
where $\mathbf{R(\phi})$ can be any appropriate correlation matrix.
When we account for the autocorrelation using the AR(1) correlation
matrix (Eq. 10), we obtain $\hat{\beta}_{1}=-1.10$ and a 95\% confidence
interval of $[-2.61,0.41${]}. The 95\% confidence interval now covers
zero and is approximately twice as wide compared to the model that
does not account for autocorrelation. The fitted trend lines for the
two models ($\hat{\beta}_{0}+\hat{\beta}_{1}\mathbf{t}$) appear nearly
identical, but when the temporal process $\boldsymbol{\eta}$ is included
($\hat{\beta}_{0}+\hat{\beta}_{1}\mathbf{t}+\hat{\boldsymbol{\eta}}$;
where $\boldsymbol{\eta}$ is estimated using Eq. \ref{eq:spatial3})
the fit is much better because the residuals appear to be uncorrelated
(see Fig. D1-D4 in Appendix D). 

To demonstrate two different first-order model specifications that
account for temporal autocorrelation, we use eigen basis vectors obtained
from a spectral decomposition of the the AR(1) correlation matrix
(Fig. 3b), as well as compactly supported uniform kernel basis function
with knots placed at each year (Fig. 3c; Appendix D). In contrast
to the spatial model used in the bioluminescence example, the AR(1)
correlation matrix does not have a corresponding kernel that can be
used as an approximation. Consequently, the first-order model that
uses a uniform kernel basis function results in a different fit to
the data when compared to the models that used an AR(1) correlation
matrix (cf., Fig. 3b to 3c). Both models, however, appear to capture
the temporal autocorrelation and result in similar estimates ($\hat{\beta}_{1}=-1.28$,
95\% confidence interval $[\textnormal{-2.59, 0.03]}$ for the uniform
kernel basis function model).

This example also demonstrates that it is important to check for multicollinearity
between basis vectors and covariates when modeling autocorrelation.
As with traditional regression models, severe multicollinearity can
negatively influence inference (\citealt{dormann2013collinearity}).
The potential for multicollinearity is evident in the first-order
specification, however, the collinearity is effectively ``hidden''
in the correlation matrix of the second-order specification (\citealt{Hodges2010,Hanks2015}).
For example, the coefficient of determination between the covariate
year ($\mathbf{t}$ in Eq. 18) and second eigen basis vector is $R^{2}=0.80$.

\subsection*{\textmd{\textsc{\normalsize{}Example 3: Predicting the Distribution
of a Species}}}

In this example, we fit three different models that account for spatial
autocorrelation to illustrate concepts presented in previous sections.
Many ecological studies aim to predict the presence or abundance of
a species at unsampled locations using species distribution models
applied to count, presence-absence, and presence-only data (\citealt{elith2009species}).
Generalized linear mixed models with a spatial random effect are well-suited to
model a species distribution using count or presence-absence data
(\citealt{Bolker2009}). For example, \citet{hooten2003predicting}
used a binary spatial regression model to predict the probability
of pointed-leaved tick trefoil (\textit{Desmodium glutinosum}) occurring
in $10\times10$ m plots across a 328 ha area from presence-absence
data collected at 216 plots (Fig. 4a). A common problem when predicting
the distribution of a species is that data are sparse relative to
the prediction domain. For this example, only 0.66\% of the plots
within the prediction domain were sampled (Fig. 4a). In this application,
\citet{hooten2003predicting} specified a second-order spatial random
effect to increase the predictive ability of a binary regression model
and to account for spatial autocorrelation generated by a complex
ecological process. A suitable second-order spatial binary model for
presence-absence data is
\begin{equation}
\begin{aligned}\mathbf{y} & \sim\text{Bernoulli}\left(g(\mathbf{X\boldsymbol{\beta}}+\boldsymbol{\eta})\right)\\
\boldsymbol{\eta} & \sim\text{N\ensuremath{\left(\mathbf{0},\mathbf{\sigma_{\alpha}^{\text{2}}R}(\phi)\right)}}\:,
\end{aligned}
\label{eq: spatial1}
\end{equation}
where $\mathbf{y}$ is an $n\times1$ vector with elements equal to
$1$ if the species is present and $0$ if the species is absent at
a sampled location, $g(\cdot)$ is the probit link function, and $\boldsymbol{\eta}$
is the spatial random effect. As in \citet{hooten2003predicting},
we specified the correlation matrix in Eq. 19 using an exponential
correlation function 
\begin{equation}
r_{ij}(\phi)=e^{-\frac{d_{ij}}{\phi}}\:,\label{eq:spatial2-2}
\end{equation}

\noindent where $d_{ij}$ is the distance between locations $i$ and
$j$. Although there are numerous ways to implement the binary regression
model, we adopt a Bayesian approach. The associated prior distributions
and covariates are described in \citet{hooten2003predicting}. After
fitting the model, we predicted the probability of occurrence at all
32,768 plots within the prediction domain. The predicted probability
of occurrence depends on several covariates and has a strong spatial
component (Fig. 4b). Unlike the Gaussian kernel basis function which
approximates a Gaussian correlation function (see bioluminescence
example), there is no kernel basis function that can approximate an
exponential covariance function (see Fig. 2 in \citealp{Higdon2002}
for details). 

Evaluating the likelihood for any second-order model requires inverting
the correlation matrix $\mathbf{R}(\phi)$. For the geostatistical
(continuous space) spatial model, inverting the correlation matrix
has a computational cost that increases according to the cube of the
sample size. For this example, when $n=216$, fitting the Bayesian
spatial model requires approximately 45 seconds per 1,000 Markov chain
Monte Carlo (MCMC) samples obtained on a laptop computer with a 2.8
GHz quad-core processor, 16 GB of RAM, and optimized basic linear
algebra subprograms, but would require about an hour per 1,000 MCMC
samples obtained from the same model if the sample size was $n=1,000$.
For large spatial datasets, a variety of dimension reduction and computationally
efficient implementations can be used to model the spatial autocorrelation. The
majority of efficient methods involve modeling the spatial autocorrelation
using basis functions and a first-order model specification. To illustrate
dimension reduction, we model the spatial autocorrelation using two
different types of basis functions: the predictive process and thin
plate regression splines. The predictive process is similar to kernel
regression methods, except the basis expansion is slightly different
and the basis coefficients are correlated in geographic space (\citealt{Banerjee2008,banerjee2014hierarchical}).
The predictive process approach models the spatial process by smoothing
over a finite number of representative locations as follows
\begin{equation}
\begin{aligned}\mathbf{Z}(\phi) & \equiv\mathbf{C\text{(}\phi\text{)}R^{*}}(\phi)^{-1}\\
\boldsymbol{\alpha} & \sim\text{N\ensuremath{\mathbf{\left(0,\mathbf{\sigma_{\alpha}^{\text{2}}R^{*}}(\phi)\right)}}}\:,
\end{aligned}
\label{eq:spatial3-1}
\end{equation}
where $\mathbf{R^{*}}(\phi)$ is the $m\times m$ correlation matrix
for preselected knots (Fig. 4c) and $\mathbf{C}(\phi)$ is the $n\times m$
correlation matrix between the observed data and knots. Using the
predictive process method with $m=50$ knots, the Bayesian model requires
approximately three seconds per 1,000 MCMC samples obtained and the
predicted probability of occurrence looks similar when compared to
the second-order spatial model (cf. Fig. 4b to Fig. 4c; Appendix E).
Furthermore, the predictive process method can be implemented using
readily available software (\citealt{finley2013spbayes}).

Generalized additive models (GAMs) are similar to models that use
spatial random effects, but rely on techniques and basis functions
commonly used in semiparametric regression (\citealt{Ruppert2003}).
Specifying GAM typically require choosing a type of basis function,
the number of basis vectors, and the location and number of knots.
The most significant difference between the previous methods we have
demonstrated and GAMs is the type of basis functions used. Many different
basis functions are used to specify GAMs and introductions can be
found in \citet{Hastie2009}, \citet{James2013}, \citet{Ruppert2003},
and \citet{Wood2006}. Although GAMs are implemented under a Bayesian
paradigm (\citealt{crainiceanu2005bayesian,Gelman2013}; Ch. 20),
penalized maximum likelihood methods are commonly used (\citealt{Wood2006}).
For illustrative purposes, we implement a GAM using thin plate regression
splines to model the spatial autocorrelation. This implementation
is available in standard software and may be particularly useful for
very large data sets (e.g., $n\approx10^{6})$, requiring approximately
two seconds to fit the model to our data using 50 basis coefficients
(\citealt{Wood2015}). The predicted probability of occurrence is
shown in Fig. 4d and is comparable to both specifications of the Bayesian
spatial model (Fig. 4). We expected similarity between the GAM and
the spatial model because there is a connection between first-order
models that use spline basis functions and second-order spatial models
(\citealt{nychka2000spatial}).

\subsection*{\textmd{\textsc{\normalsize{}Discussion}}}

\subsubsection*{\textmd{\textit{Autocorrelation: the two cultures}}}

``What is one person's covariance structure is another persons's
mean structure (\citealt{cressie1993statistics}; pg. 25).'' Within
subfields of statistics that focus on dependent data (e.g., spatial
statistics), there is no general consensus on whether the influence
of autocorrelation should be specified in the mean or covariance structure
of a probability distribution. The notion that first-order specified
models that use basis functions and second-order specified spatial
and time-series models are both useful for dependent data has been
a topic of discussion for several decades among statisticians (e.g.,
\citealt{Cressie1989} and comments by \citealt{Wahba1990}; \citealt{Laslett1994}
and comments by \citealt{Handcock1994} and \citealt{Laslett1994comment2}).
As we have demonstrated, the two approaches can result in the same
model or an approximation thereof. With regard to which method to
use, there are entire books written about correlation functions from
a single perspective (e.g., Kriging in a spatial context; \citealt{stein2012interpolation})
and about certain classes of basis functions (\citealt{nason2010wavelet};
Table 2). Given the diversity of approaches, it is difficult to make
specific recommendations. Our goal is to encourage researchers to
consider both perspectives, rather than one or the other.

\subsubsection*{\textmd{\textit{First-order or second-order?}}}

Models that use second-order specifications can be converted to the
equivalent first-order specification to assess collinearity among
basis vectors and covariates of interest (e.g., \citealt{Hodges2010}).
Modeling the autocorrelation using a first-order specification can
be beneficial when the autocorrelation does not follow a standard
correlation function, such as the case with data collected from streams
and rivers (\citealt{peterson2010mixed,sampsonnonstationarykernel,isaak2014applications})
or moving animals (\citealt{buderman2015functional}). Although we
have not demonstrated this in our examples, the first-order specification
might be more appealing when specifying theory-based ecological models
(e.g., using partial differential equations), because the first-order
model specification is naturally hierarchical (\citealt{wikle2010general}).
Using the first-order specification, the conditional distribution
of the data (or unobserved latent process) can be selected and knowledge
of the process can be incorporated into the mean structure (e.g.,
\citealt{Wikle2003,Hooten2008,wikle2010general}). Although scientifically
meaningful, second-order model specifications may be more challenging
to understand when compared to first-order models specifications (\citeauthor{hanks2017constructive}
\textit{under review}). Second-order specifications, however, can
facilitate more computationally efficient algorithms. Thus, many contemporary
models for autocorrelation are specified in terms of first-order structure
and then converted to second-order structure for implementation (e.g.,
\citealt{finley2013spbayes}).

\subsubsection*{\textmd{\textit{Choosing basis functions}}}

Choosing basis functions requires an understanding of both the underlying
ecological process and the properties of the basis functions. For
example, a property of the polynomial basis function is that it has
a global support, thus an observation at one location influences the
fit of the model at another location, no matter how far apart the
two locations are. This is why polynomial basis expansions often fails
to model fine scale structure (cf., Fig. 1b to 2b). From an ecological
perspective, the global support of polynomial basis functions implies
that the underlying ecological process is connected across the entire
space of interest. If the ecological process is thought to have discontinues,
then basis functions that capture discontinuous structure and have
compact support are a better choice (e.g., the uniform kernel used
in Example 2; Table 2). 

When selecting a basis function to model autocorrelation, standard
model checking procedures are critical to ensure that model assumptions
are met (e.g., checking for correlated residuals, multicollinearity,
lack of fit, overfitting). Formal model selection may also be useful
for selecting the optimal basis functions (\citealt{gelfandknotdesign};
\citealt{Gelman2013} Ch. 20; \citealt{Hooten2014}). Computational
considerations may also be important when choosing a basis function.
For example, orthogonal basis functions often result in more stable
computational algorithms because the basis vectors are independent,
obviating collinearity between basis vectors. We illustrated only
a small fraction of the basis functions that could be used, thus we
recommend that practitioners become familiar with the variety of options
to ensure that the chosen basis function matches the goals of the
study. To facilitate this, we have provided a brief summary of common
basis functions, their properties, and useful references in Table
2.

\subsubsection*{\textmd{\textit{Implementation}}}

Typically, only a small number of covariates are included in a regression
model, but one may want to include as many or more basis vectors as
there are observations. For example, there are as many eigen basis
vectors as there are unique locations in the dataset when the correlation
matrix is specified using a correlation function. When many basis
vectors are used to model autocorrelation, the model can overfit the
data. Adding constraints to high-dimensional estimation problems is
a common technique to prevent overfitting. Such methods include regularization,
penalized maximum likelihood estimation (e.g., ridge regression),
treating the basis coefficients as random effects, or using a prior
distribution that induces shrinkage (regularization) in a Bayesian
model. There are important connections between methods that impose
constraints to prevent overfitting that we have not presented here,
but are important to understand when implementing models that use
basis functions (\citealt{Hooten2014}).

When fitting models data sets where dimension reduction is required,
there is a trade-off between the reduction in dimension and the fit
of the model. The fit of the model is influenced by dimension reduction
because choosing the number of basis vectors to include in a model
is an implicit form of regularization (\citealt{Gelman2013}, Ch.
20; \citealt{Hooten2014}). Determining which basis functions are
optimal for approximating correlation functions, how many basis vectors
are needed, and the locations of knots are active areas of research
\citep{gelfandknotdesign}. A general rule of thumb is to choose fewer
basis vectors than the number of unique locations in the dataset,
but as large as possible given the computational restrictions so that
the predictions are accurate (e.g., on out-of-sample data). A detailed
summary of dimension reduction approaches is beyond the scope of our
work, but accessible introductions can be found in \citet{paciorek2007computational},
\citet{wikledimred}, \citet{cressie2011statistics}, and \citet{banerjee2014hierarchical}.

Basis function model specifications have also become popular in spatio-temporal
modeling, both in environmental (e.g., \citealt{wikle2002kernel})
and ecological applications (e.g., \citealt{Hooten2007}). In practice,
we find that understanding the properties of basis functions is critical
to implementing computationally efficient Bayesian hierarchical models
that account for spatial, temporal, or spatio-temporal autocorrelation.
In addition, using basis functions as part of a Bayesian hierarchical
model makes many spatio-temporal models accessible to users of JAGS,
NIMBLE, Stan, and WinBugs (\citealt{crainiceanu2005bayesian}). Using
the tools and techniques we have presented in this paper, basis function
components can be added to existing hierarchical models to account
for autocorrelation.

\subsubsection*{\textmd{\textit{Inference and collinearity}}}

For some applications, collinearity among covariates and basis vectors
might occur and the development of remedial methods is a current topic
of research in spatial statistics (\citealt{Reich2006,paciorek2010importance,Hodges2010,hodges2013richly,hughes2013dimension,Hanks2015}).
The effects of collinearity among covariates and basis vectors have
been noted in the ecological literature as well, particularly in a
spatial context (e.g., \citealt{Bini2009,kuhn2007incorporating}).
In our experience, collinearity among covariates and basis vectors
is a difficult challenge in applied problems. In some cases, the conventional
wisdom that applies to collinearity among covariates can also be applied
to basis vectors, but new intuition is needed when basis coefficients
are treated as random effects (\citealt{Hodges2010,paciorek2010importance,hodges2013richly,Hanks2015,murakami2015random}).
As with collinearity among covariates in linear regression models,
there is no clear remedy for extreme cases.

\subsubsection*{\textmd{\textit{Conclusion}}}

Ecologists face many choices when specifying models. One important
choice is how to model autocorrelation. Autocorrelation, however,
is not limited to specific domains and can occur in any space (e.g.,
covariate space, time, three-dimensional Euclidian space). Many methods
used to model autocorrelation are general and can be understood as generalized
linear mixed models that employ basis expansions and treat basis coefficients
as random effects. Using the basis function approach, we find that
many of the commonly used ecological models can be modified to incorporate
autocorrelation.

\subsection*{\textmd{\textsc{\normalsize{}Acknowledgements}}}

We thank Evan Cooch, Perry de Valpine, Devin Johnson, and three anonymous
reviewers for valuable insight and early discussions about this work.
The authors acknowledge support for this research from USGS G14AC00366.
Any use of trade, firm, or product names is for descriptive purposes
only and does not imply endorsement by the U.S. Government.

\vspace{-0.3in}
\renewcommand\refname{\textmd{\textsc{\normalsize{}Literature Cited}}} 

\setlength{\bibsep}{0pt}

\bibliographystyle{apa}
\bibliography{ref_basis_functions_ms}

\subsection*{\textmd{\textsc{\normalsize{}Supplemental Material}}}

\textbf{Appendix A }

\noindent Additional comments about basis functions.

\noindent \textbf{Appendix B}

\noindent Interactive illustration of Fig. 1 \& Fig. 2 https://hootenlab.shinyapps.io/hooten\_lab\_manuscript\_figures.

\noindent \textbf{Appendix C}

\noindent Tutorial with data and R code for the bioluminescence example.

\noindent \textbf{Appendix D}

\noindent Tutorial with data and R code for the time series example.

\noindent \textbf{Appendix E}

\noindent Tutorial with data and R code for the spatial example.

\pagebreak{}

\begin{landscape}

\noindent Table 1. Glossary of terms and definitions.

\begin{singlespace}
\begin{tabular}{ll}
\hline 
Term & Definition\tabularnewline
\hline 
\multirow{1}{5cm}{Autocorrelation} & \multirow{2}{18cm}{Correlation between observations based on some measure of distance or time that exists after the influence of all covariates is accounted for}\tabularnewline
 & \tabularnewline
\multirow{1}{5cm}{Basis expansion} & \multirow{1}{*}{\multirow{1}{18cm}{A collection of basis vectors from a single covariate}}\tabularnewline
\multirow{1}{5cm}{Basis vector} & \multirow{1}{18cm}{Any transformation of a covariate}\tabularnewline
\multirow{1}{5cm}{Basis function} & \multirow{1}{18cm}{Any mathematical function that transforms a covariate}\tabularnewline
\multirow{1}{5cm}{Compact support} & \multirow{1}{18cm}{A support that does not includes all possible locations or time points}\tabularnewline
\multirow{1}{5cm}{Correlation function} & \multirow{1}{18cm}{A function that describes the autocorrelation between observations}\tabularnewline
\multirow{1}{5cm}{Correlation matrix} & \multirow{1}{18cm}{A positive semi-definite matrix whose elements are the correlation between observations}\tabularnewline
\multirow{1}{5cm}{Covariate} & \multirow{1}{18cm}{Any quantity that can be measured and is associated with an observation (e.g., the time or spatial location of the observation)}\tabularnewline
 & \tabularnewline
\multirow{1}{5cm}{Dependence} & \multirow{2}{18cm}{Correlation between observations defined in a general space (spatial or temporal dependence is equivalent to autocorrelation) }\tabularnewline
 & \tabularnewline
\multirow{1}{5cm}{First-order specification} & \multirow{1}{18cm}{When a function that models the dependence is specified in the mean (expected value) of a probability distribution}\tabularnewline
 & \tabularnewline
\multirow{1}{5cm}{Global support} & \multirow{1}{18cm}{A support that includes all possible locations or time points}\tabularnewline
\multirow{1}{5cm}{Second-order specification} & \multirow{1}{18cm}{When a function that models dependence is specified in the covariance of a probability distribution}\tabularnewline
\multirow{1}{5cm}{Support} & \multirow{1}{18cm}{The set of locations or time points where the basis function results in non-zero values}\tabularnewline
\hline 
\end{tabular}\pagebreak{}

Table 2. Common types of basis functions, important properties and
references.

\begin{tabular*}{1.2\textwidth}{@{\extracolsep{\fill}}>{\raggedright}p{0.17\textwidth}>{\raggedright}p{0.14\textwidth}>{\raggedright}p{0.16\textwidth}>{\raggedright}p{0.3\textwidth}>{\raggedright}p{0.38\textwidth}}
\hline 
Basis function & Orthogonal & Support & Notable use & Reference\tabularnewline
\hline 
Eigen & yes & global & Dimension reduction and detecting multicollinearity between basis
vectors and covariates in second-order models & \citet{Hodges2010}, \citeauthor{hodges2013richly} (2013; Ch. 10),
\citeauthor{cressie2011statistics} (2011; Ch. 5)\tabularnewline
Fourier & yes & global & Large data sets with a smooth effect of autocorrelation & \citet{paciorek2007bayesian}, \citeauthor{cressie2011statistics}
(2011; Ch. 3)\tabularnewline
Kernel & no & global or compact & Large data sets and a directional effect of autocorrelation & Higdon (2002), Sampson (2010), Peterson and Ver Hoef (2010)\tabularnewline
Piecewise linear & no & local & Implementing numerical solutions to stochastic partial differential
equations & \citet{lindgren2011explicit}, \citet{INLASPDE}\tabularnewline
Polynomial & no & global & Modeling simple nonlinear effects of autocorrelation & Carroll et al. (2003; Ch. 2), James et al. (2013; Ch. 7)\tabularnewline
Splines & no & global or compact & Large data sets with smooth effects of autocorrelation & Carroll et al. (2003; Ch. 3), Wood et al. (2006), Hastie et al. (2009;
Ch. 5), James et al. (2013; Ch. 7)\tabularnewline
Wavelets & yes & global and compact & Modeling discontinuous effects of autocorrelation & Nason (2008)\tabularnewline
\hline 
\end{tabular*}
\end{singlespace}

\end{landscape}

\pagebreak{}

\noindent \textbf{Figure 1.} Scatterplots showing the density of pelagic
bioluminescence (sources) versus water depth (points). The top panels
show fitted regression models (black lines). The corresponding basis
vectors multiplied by the estimated coefficients (colored curves)
are shown in the bottom panels. Simple linear regression model (a;
Eq. 1) with corresponding constant (red) and linear basis vectors
(blue). Polynomial regression model (b; Eq. 2) with corresponding
constant (red), linear (blue), and quadratic basis vectors (green).
Polynomial regression model (c) similar to that shown in panel (b),
except with basis vectors calculated relative to three water depths
($k_{j}$ in Eq. 3; vertical colored lines). Note that the basis vectors
are multiplied by the estimated coefficients and summed to produce
the fitted curves (black lines). See Appendix B for an interactive
version of this figure.

\noindent \textbf{Figure 2.} Scatterplots showing the density of pelagic
bioluminescence (sources) versus water depth (points). The top panels
shows the fitted curve (black lines) obtained from a second-order
specification that uses a Gaussian correlation function (a), the equivalent
first-order specification that uses eigen basis vectors (b), and a
first-order specification that uses a Gaussian kernel basis function
(c). The bottom panels show the intercept term (red), eigen basis
vectors (b), and Gaussian kernel basis vectors (c). For illustrative
purposes, only the product of three basis vectors and coefficients
are shown (with knots located at the vertical lines in panel (c)).
There are 51 eigen basis vectors and coefficients that sum to produce
the fitted curve (black line) in panel (b), and 17 kernel basis vectors
that sum to produce the fitted curve (black line) in panel (c). See
Appendix B for an interactive version of this figure.

\noindent \textbf{Figure 3.} Scatterplots of bobwhite quail population
indices over time (points). The top panels show fitted regression
models (black lines) obtained from a second-order specification that
uses an AR(1) correlation matrix (a), the equivalent first-order specification
that uses eigen basis vectors (b), and a first-order specification
that uses a compactly supported uniform kernel basis function (c). The
bottom panels show the fixed effects term (red), three eigen basis
vectors (b), and three compactly supported kernel basis vectors with
knots located at the vertical lines (c). All basis vectors are multiplied
by basis coefficients.

\noindent \textbf{Figure 4.} Prediction domain from the Missouri Ozark
Forest Ecosystem Project presented in \citet{hooten2003predicting}.
Red and black points (a) represent the $10\times10$ m plot locations
that were sampled ($n=216)$ and whether pointed-leaved tick trefoil
was present (red) or absent (black). The heat maps (b,c,d) show the
predicted probability of occurrence in 32,768 plots from a binary
spatial regression model (b; Eq. 19), a reduced dimension binary spatial
regression model using predictive process basis functions (Eq. 21)
with knots located within the prediction domain represented by + (c),
and a generalized additive model that uses thin plate regression splines
(d). 

\pagebreak{}

\begin{landscape}

\begin{center}
\begin{figure}[H]
\protect\caption{\protect\includegraphics{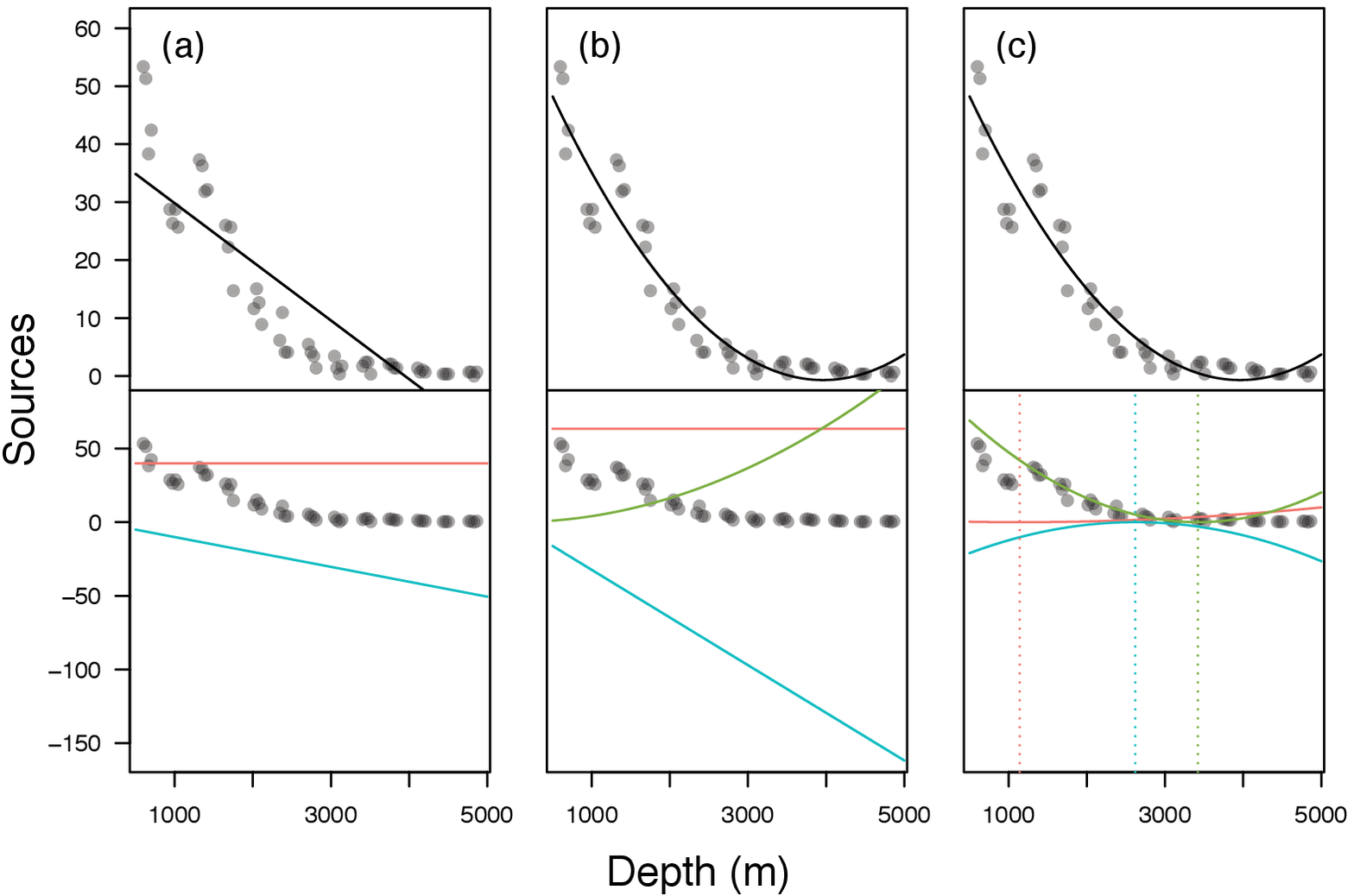}}
\end{figure}

\par\end{center}

\end{landscape}

\pagebreak{}

\begin{landscape}

\begin{center}
\begin{figure}[H]
\protect\caption{\protect\includegraphics{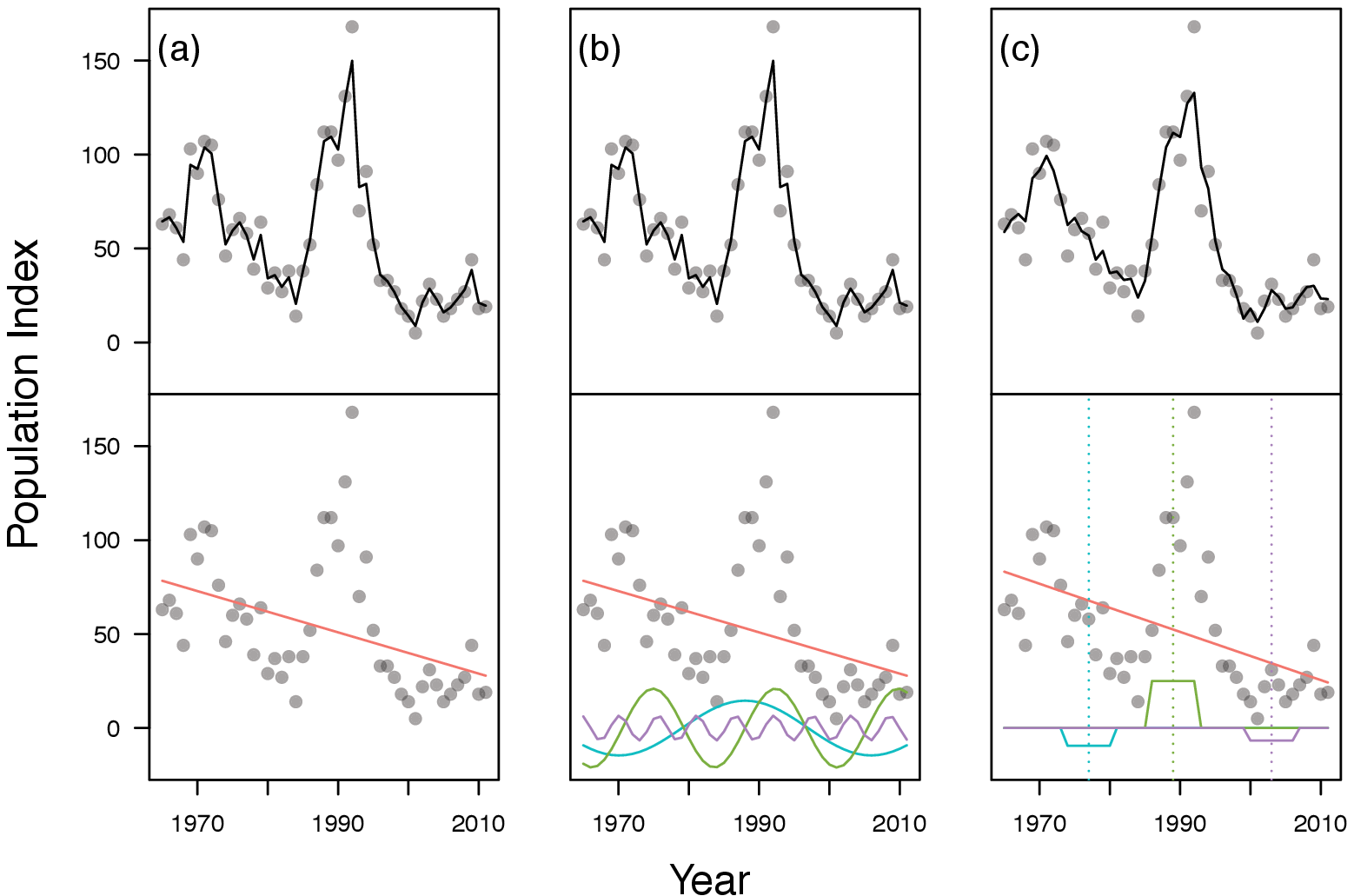}}
\end{figure}

\par\end{center}

\end{landscape}

\pagebreak{} \begin{landscape}

\begin{center}
\begin{figure}[H]
\protect\caption{\protect\includegraphics{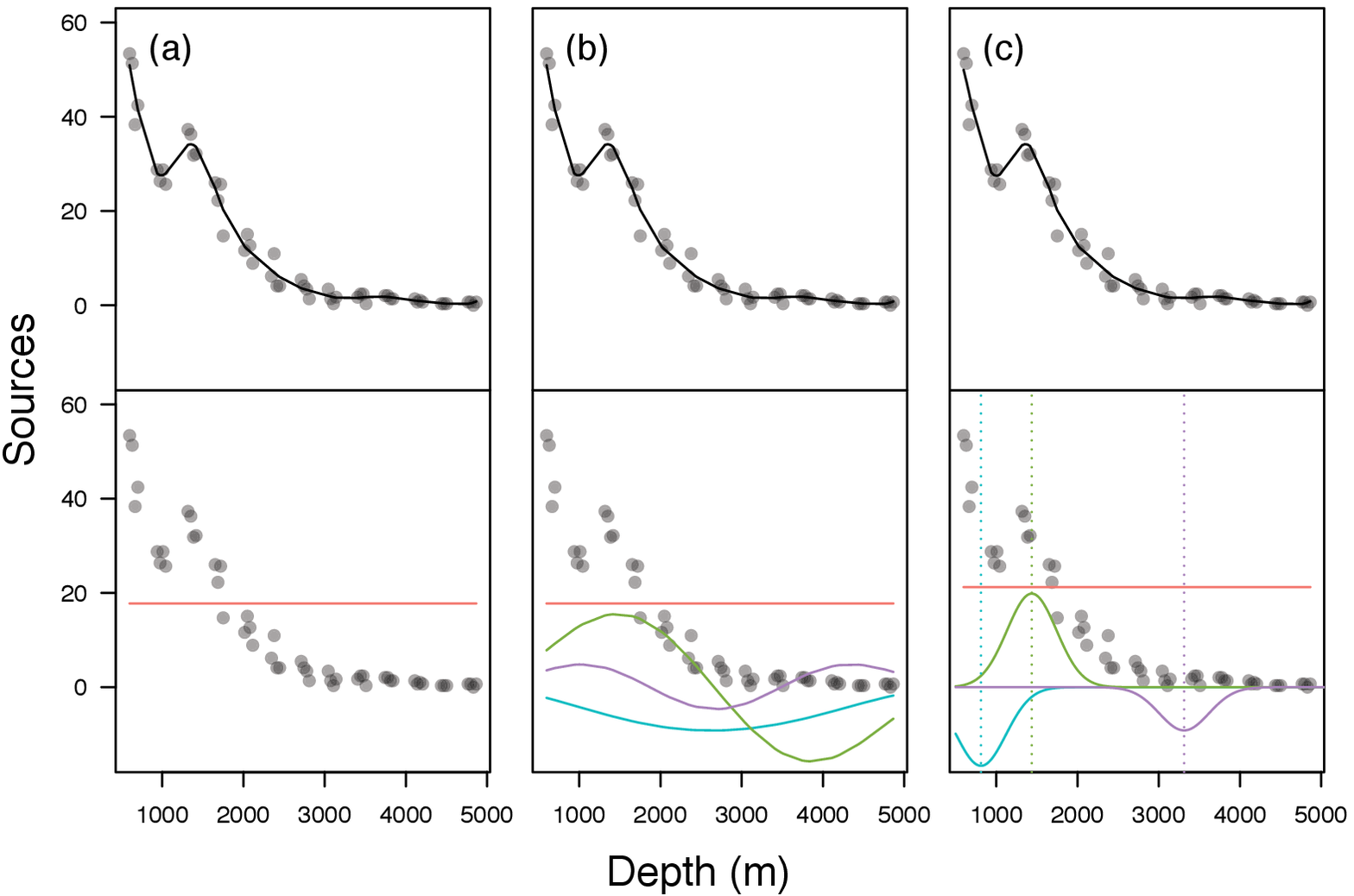}}
\end{figure}

\par\end{center}

\end{landscape}

\pagebreak{}

\begin{center}
\begin{figure}[H]
\protect\caption{\protect\includegraphics{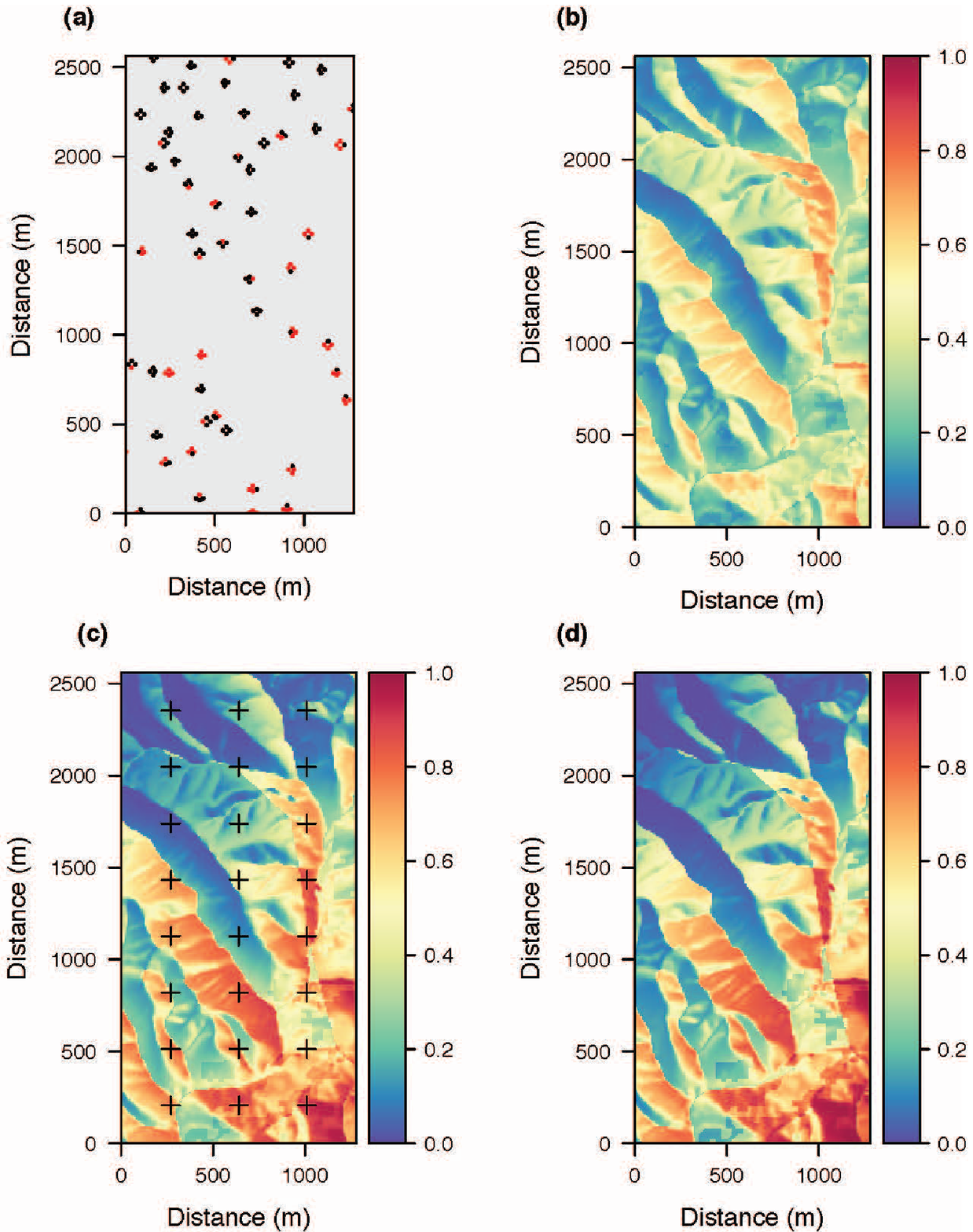}}
\end{figure}

\par\end{center}

\end{spacing}
\end{flushleft}
\end{document}